\documentclass[pra, aps, twocolumn, floatfix, showpacs]{revtex4-1}
\usepackage{graphicx, amsmath, amssymb, times, subfigure}

\begin{document}
\title{Topological superfluids on a square optical lattice with non-Abelian gauge fields: \\
Effects of next-nearest-neighbor hopping in the BCS-BEC evolution}
\author{M. Iskin}
\affiliation{
Department of Physics, Ko\c c University, Rumelifeneri Yolu, 34450 Sar{\i}yer, Istanbul, Turkey.
}
\date{\today}
\begin{abstract}

We consider a two-component Fermi gas with attractive interactions on a square 
optical lattice, and study the interplay of Zeeman field, spin-orbit coupling 
and next-nearest-neighbor hopping on the ground-state phase diagrams in the 
entire BCS-BEC evolution. 
In particular, we first classify and distinguish all possible superfluid phases 
by the momentum-space topology of their zero-energy quasiparticle/quasihole 
excitations, and then numerically establish a plethora of quantum phase 
transitions in between. These transitions are further signalled and evidenced 
by the changes in the corresponding topological invariant of the system, 
\textit{i.e.}, its Chern number. Lastly, we find that the superfluid phase exhibits 
a reentrant structure, separated by a fingering normal phase, the origin of which 
is traced back to the changes in the single-particle density of states.

\end{abstract}
\maketitle

\section{Introduction}
\label{sec:intro}

In an archetypal square optical lattice~\cite{esslinger10}, since either the lattice 
potential is separable in $x$ and $y$ directions, and therefore, the atoms are 
strictly not allowed to tunnel (hop) along the diagonal, \textit{i.e}, the next-nearest 
neighbor (NNN), directions, or the diagonal hoppings are negligible, 
the single-particle dynamics is controlled by the nearest-neighbor (NN) hoppings. 
However, it has been recently argued that the ratio of NNN to NN hopping can 
be effectively tuned all the way from $0$ to $\infty$ in driven optical lattices, 
by periodically shaking the lattice potential in time, \textit{e.g.}, 
see ~\cite{beugeling12, miao15}. Indeed, pioneering experiments on such lattices 
have already paved the way towards `tunnelling engineering', allowing for 
tunable (real and/or complex) NN and NNN hoppings~\cite{lignier07, struck12, parker13}. 
For instance, applying similar techniques to a honeycomb optical lattice, 
and by controlling not only the amplitude but also the phase of the 
complex NNN tunnelling~\cite{jotzu14}, the ETH Zurich group has recently 
reported the very first realization of the the topological Haldane model~\cite{haldane88} 
with ultracold fermions. 

In addition to the driven optical lattices, other methods relying on the laser-assisted 
tunnelling and potential energy gradient have also been used to engineer 
spatially-dependent complex hoppings, by coupling the internal/atomic and 
external/space degrees of freedom, which are analogous to having artificial 
gauge fields~\cite{dalibard11, galitski13}. The recent realizations of the 
Hofstadter-Harper model are some of the prime 
examples~\cite{aidelsburger13, miyake13}. Indeed, the fruitful quest for
creating various Abelian or non-Abelian artificial gauge fields in discrete/lattice 
and continuum systems has been pursued by many groups in the past 
decade or so, and there are numerous ultracold realizations including the spin-orbit 
coupling (SOC)~\cite{chen12, wang12, cheuk12, qu13a, fu13, williams13}.

Motivated by the advent of driven optical lattices with tunable hoppings, here we 
consider a two-component Fermi gas with attractive interactions on a square optical 
lattice, and study the interplay of Zeeman field, SOC and NNN hopping 
on the ground-state phase diagrams in the entire BCS-BEC evolution. 
In particular, we firmly establish a great deal of quantum phase transitions 
between superfluid (SF) phases with distinct momentum ($\mathbf{k}$) space 
topologies, that are driven mainly 
by the Zeeman field. In addition, we also derive analytical expressions for the 
changes in the Chern number (CN) showing that these phase transitions are further 
signalled and evidenced by the changes in the topological invariant of the system 
when there is SOC. Furthermore, we find that the SF phase exhibits a reentrant 
structure, separated by a fingering normal phase, as a function of total particle
filling. This intricate structure is a result of combined SOC and 
NNN hopping, and we trace its origin back to the single-particle density of 
states (DoS) of the system which is shown to exhibit a number of narrow strips 
(depending on the SOC strength) as a function of energy. 

The rest of the paper is organized as follows. In Sec.~\ref{sec:mft}, 
we first introduce the model Hamiltonian, then discuss the effects of SOC both on the 
single-particle problem and on the corresponding DoS, and then 
derive the self-consistency mean-field equations for handling the many-body 
problem. In Sec.~\ref{sec:topo}, we first classify and distinguish all possible SF 
phases by the $\mathbf{k}$-space topology of their zero-energy quasiparticle/quasihole 
excitations, and then derive the corresponding changes in the CN of the system. 
Our numerical calculations and the resultant ground-state phase diagrams are 
presented and thoroughly analyzed in Sec.~\ref{sec:numerics}, and the paper 
ends with a brief summary of our conclusions and an outlook in Sec.~\ref{sec:conc}.

\section{Mean-Field Theory}
\label{sec:mft}

To explore the possible quantum phases of spin-orbit coupled Fermi gases 
loaded into the periodic arrays of an optical lattice potential, we limit our 
discussion to the tight-binding atom-hopping regime. The single-band 
Hubbard model description is known to serve quite well in this regime,
for which the gauge fields can be taken into account via the Peierls substitution. 
In particular, we study the topological effects of a generic 
non-Abelian gauge field~\cite{kubasiak10}
\begin{equation}
\mathbf{A}=(\alpha \sigma_y, -\beta\sigma_x)
\label{eqn:gauge}
\end{equation}
on two-dimensional lattice Fermi gases, where $\sigma_x$ and $\sigma_y$ 
are the Pauli-spin matrices, and the parameters $\alpha \ge 0$ and 
$\beta \ge 0$ characterize both the strength and the symmetry of the SOC. 
Since the atom-atom interactions are extremely short-ranged in a typical 
cold-atom setting, the interactions are assumed to be onsite and attractive, 
and we treat such an interaction term in the Hamiltonian with the BCS 
mean-field approach for the entire BCS-BEC evolution. It is well-accepted 
in the cold-atom community that similar treatments for related problems 
have provided qualitatively correct descriptions of atomic systems at 
least for their ground states.

\subsection{Model Hamiltonian}
\label{sec:ham}

As a result, we study ground-state phases of the following Hamiltonian
\begin{align}
\label{eqn:ham}
H &= - \sum_{\sigma \sigma' ij} c_{\sigma' j}^\dagger t_{j i}^{\sigma' \sigma} c_{\sigma i} 
- h \sum_i \left( c_{\uparrow i}^\dagger c_{\uparrow i} - c_{\downarrow i}^\dagger c_{\downarrow i} \right) \\
& + \sum_i \left( \Delta_i c_{\uparrow i}^\dagger c_{\downarrow i}^\dagger 
+ \Delta_i^*c_{\downarrow i} c_{\uparrow i} + \frac{|\Delta_i|^2}{g} \right)
- \mu \sum_{\sigma i} c_{\sigma i}^\dagger c_{\sigma i},  \nonumber
\end{align}
where the operator $c_{\sigma i}^\dagger$ ($c_{\sigma i}$) creates (annihilates) a
pseudo-spin $\sigma = \{ \uparrow, \downarrow\}$ fermion at lattice site $i$~\cite{Wang14}. 
In the presence of a gauge field, the hopping of atoms from site $i$ to $j$ can be 
described in general by
$
t_{j i}^{\sigma' \sigma} = t_{ji} e^{-{\rm i} \theta_{ji}^{\sigma' \sigma}},
$ 
where $t_{ji}$ is its amplitude and the accumulated phase factor
$
\theta_{ji} = \int_\mathbf{r_i}^\mathbf{r_j} \mathbf{A} \cdot d\mathbf{r}
$
is a consequence of the Peierls substitution with $\mathbf{r_i}$ 
the position of site $i$. The lattice spacing $a$ is set to unity throughout 
the paper. In the pairing terms, the complex number
$
\Delta_i = g \langle c_{\uparrow i} c_{\downarrow i} \rangle
$
is the local SF order parameter for the mean-field ground state, where $g \ge 0$ 
is the strength of the onsite attraction between $\uparrow$ and $\downarrow$ 
fermions, and $\langle \cdots \rangle$ is a thermal average. Lastly, the chemical 
potential $\mu$ determines the total number $N = N_\uparrow + N_\downarrow$ 
of atoms where $N_\sigma = \sum_i n_{\sigma i}$ with the local fermion filling
$
0 \le n_{\sigma i} = \langle c_{\sigma i}^\dagger c_{\sigma i} \rangle \le 1,
$
and the out-of-plane Zeeman field $h \ge 0$ determines the polarization 
$P = (N_\uparrow - N_\downarrow)/N \ge 0$ of the system which is assumed 
to be positive without loosing generality.

Having a translationally-invariant lattice in mind, this Hamiltonian can be easily 
solved using the Fourier series expansion of the annihilation operator 
$
c_{\sigma i} = ( 1/\sqrt{M}) \sum_\mathbf{k} e^{{\rm i} \mathbf{k} \cdot \mathbf{r_i}} c_{\sigma \mathbf{k}}
$
in momentum space and its Hermitian conjugate, where $M \to \infty$ 
is the number of lattice sites in the system and $c_{\sigma \mathbf{k}}$ 
annihilates a $\sigma$ fermion with wave vector $\mathbf{k} = (k_x, k_y)$. 
For instance, the single-particle terms can be compactly written in this 
representation as
$
H_0 = \sum_{\sigma \sigma' \mathbf{k}} c_{\sigma' \mathbf{k}}^\dagger h_{0 \mathbf{k}}^{\sigma' \sigma} c_{\sigma \mathbf{k}},
$
where the $2 \times 2$ matrix
$
h_{0 \mathbf{k}}^{\sigma' \sigma} = \epsilon_\mathbf{k} \sigma_0 - h\sigma_z 
+ \mathbf{S}_\mathbf{k} \cdot \vec{\sigma}
$
describes the Hamiltonian dynamics.
Here, $\sigma_0$ is the identity matrix, 
$\vec{\sigma} = (\sigma_x, \sigma_y, \sigma_z)$ is a vector
of spin matrices, $\epsilon_\mathbf{k}$ is the energy dispersion and
$\mathbf{S}_\mathbf{k} = (S_\mathbf{k}^x , S_\mathbf{k}^y, 0)$ is the 
spin-momentum coupling (\textit{i.e.}, SOC). The specific forms of 
$\epsilon_\mathbf{k}$ and $\mathbf{S}_\mathbf{k}$ depend on the 
particular lattice geometry of interest.

\subsection{Square Lattice with NN and NNN Hoppings}
\label{sec:ham.sl}

In this paper, we consider a square crystal lattice with primitive unit vectors 
$\mathbf{a_1} = (1,0)$ and $\mathbf{a_2} = (0,1)$, and allow for NN as well as 
NNN hoppings and set all of the longer-ranged hoppings to zero for simplicity. 
While $t_{j i} = t \ge 0$ is assumed to be positive for NN hoppings, we vary 
both the strength and the sign of NNN hoppings $t_{j i} = t'$. 
Therefore, the $\uparrow$ and $\downarrow$ atoms gain
$
\theta_{i+\mathbf{\widehat{x}}, i} \rightarrow e^{-{\rm i} \alpha \sigma_y}
$
factors for NN hopping in the positive $\mathbf{\widehat{x}}$ direction, 
$
\theta_{i+\mathbf{\widehat{y}}, i} \rightarrow e^{{\rm i} \beta \sigma_x}
$
factors for NN hopping in the positive $\mathbf{\widehat{y}}$ direction, and
$
\theta_{i+\mathbf{\widehat{e}_\pm}, i} \rightarrow e^{-{\rm i} (\pm \alpha \sigma_y - \beta \sigma_x)}
$
factors for NNN hoppings in the diagonal 
$
\mathbf{\widehat{e}_\pm} = (\pm \mathbf{\widehat{x}} + \mathbf{\widehat{y}})/\sqrt{2}
$ 
directions. The phase factors for hoppings in the opposite
($-\mathbf{\widehat{x}}$, $-\mathbf{\widehat{y}}$ and $- \mathbf{\widehat{e}_\pm}$) 
directions are given by the corresponding Hermitian conjugates.
Using the identity 
$
e^{{\rm i} \mathbf{v} \cdot \vec{\sigma}} 
= \sigma_0 \cos v  + {\rm i} (\mathbf{\widehat{v}} \cdot \vec{\sigma}) \sin v
$
valid for any given vector $\mathbf{v} = v \mathbf{\widehat{v}}$, we obtain
\begin{align}
\label{eqn:ek.s}
\epsilon_\mathbf{k} &= - 2t \cos \alpha \cos k_x - 2t \cos \beta \cos k_y \nonumber \\ 
&\,\,\,\,\,\,\, - 4t' \cos \gamma \cos k_x \cos k_y, \\
\label{eqn:skx.s}
S_\mathbf{k}^x &= - 2t \sin \beta \sin k_y - 4t' \beta \frac{\sin \gamma} {\gamma} \cos k_x \sin k_y, \\ 
\label{eqn:sky.s}
S_\mathbf{k}^y &= 2t \sin \alpha \sin k_x + 4t' \alpha \frac{\sin \gamma} {\gamma} \sin k_x \cos k_y,
\end{align}
where $\gamma = \sqrt{\alpha^2 + \beta^2}$.
Note that the reciprocal of a square lattice is also a square lattice in 
$\mathbf{k}$ space with primitive unit vectors $\mathbf{b_1} = (2\pi, 0)$ 
and $\mathbf{b_2} = (0, 2\pi)$, and therefore, the first BZ is bounded 
by $|k_x| = \pi$ and $|k_y| = \pi$.

\subsection{Helicity Bands}
\label{sec:bands}

To isolate the effects of SOC on the single-particle problem, we set $h = 0$ and 
analyze the resultant band structures in the first BZ. 
When $h = 0$, the eigenvalues $\epsilon_\mathbf{k} \pm |S_\mathbf{k}|$ of 
the non-interacting Hamiltoian matrix $h_{0 \mathbf{k}}$ correspond, respectively, 
to the single-particle excitation energies for the positive and negative 
helicity branches. 

\begin{center}
\begin{figure}[htb]
\vskip -0.1cm
\centerline{\scalebox{0.29}{\includegraphics{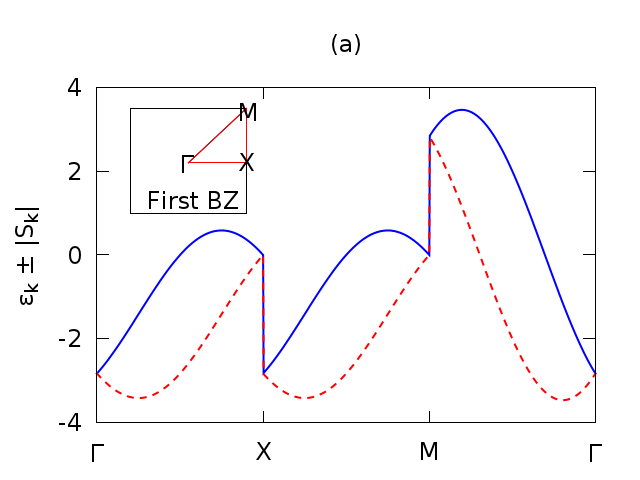}}}
\vskip -0.5cm
\centerline{\scalebox{0.29}{\includegraphics{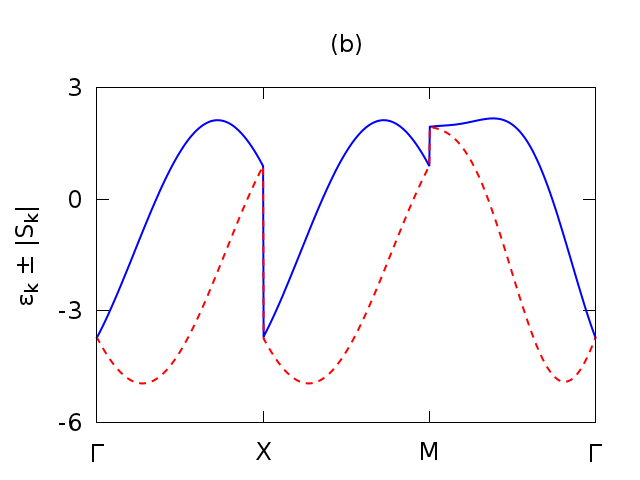}}}
\vskip -0.5cm
\centerline{\scalebox{0.29}{\includegraphics{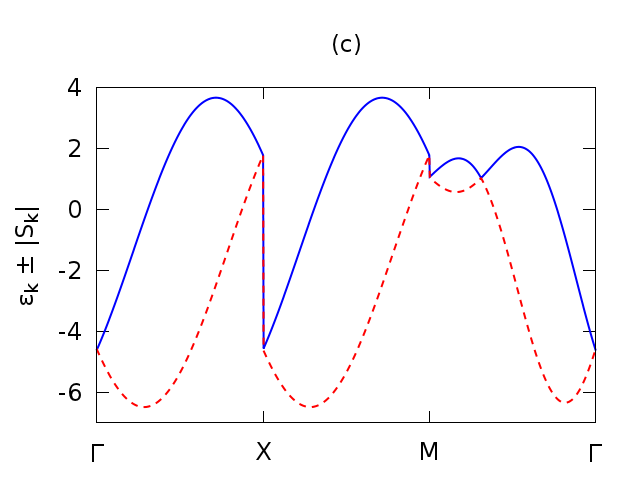}}}
\vskip -0.5cm
\centerline{\scalebox{0.29}{\includegraphics{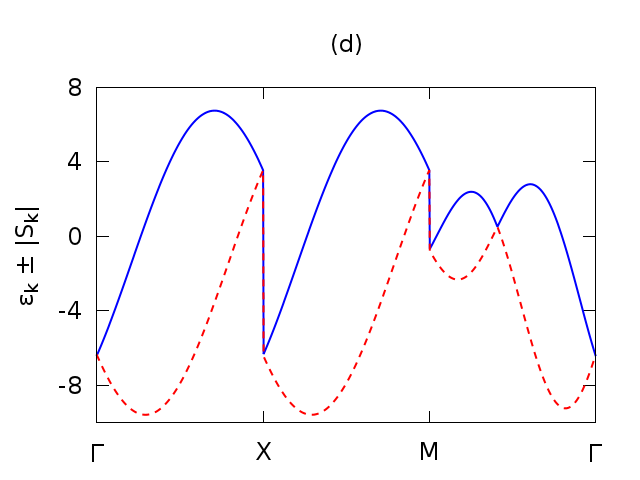}}}
\caption{\label{fig:bands} (Color online)
The positive (straight) and negative (dashed) helicity bands 
$\epsilon_\mathbf{k} \pm |S_\mathbf{k}|$ are shown in units of $t$ as a function 
of momentum $\mathbf{k}$ within the first BZ. 
Here, $\Gamma$, X and M correspond, respectively, to ($0,0$), ($\pi, 0$)
and ($\pi, \pi$) points, where (a) $t'/t = 0$, (b) $0.5$, (c) $1$ and (d) $2$,
and $\alpha = \beta = \pi/4$ in all figures.
}
\end{figure}
\end{center}

In the absence of a SOC, the positive and negative helicity bands are 
trivially degenerate (identical) everywhere in $\mathbf{k}$ space. However, 
the presence of a SOC lifts most of this degeneracy, except for five sets of 
touching points determined by the condition $|S_\mathbf{k}| = 0$.
In addition to the center $\mathbf{k_1} = (0, 0)$ of the square-shaped BZ,
the two-point set $\mathbf{k_2} = (0, \pm \pi)$ corresponds to the midpoints of
the top and bottom edges of the BZ adding in total to a one full point, 
the two-point set $\mathbf{k_3} = (\pm \pi, 0)$ corresponds to the midpoints of
the right and left edges of the BZ adding in total to a one full point, 
the four-point set $\mathbf{k_4} = (\pm \pi, \pm \pi)$ corresponds to the four 
corners of the BZ adding again to a one full point, and
the four-point set $\mathbf{k_5} = (k_5^x, k_5^y)$ is such that
$
\cos k_5^x = - \frac{\sin \beta}{2\beta} \frac{\gamma}{\sin \gamma} \frac{t}{t'}
$
and
$
\cos k_5^y = - \frac{\sin \alpha}{2\alpha} \frac{\gamma}{\sin \gamma} \frac{t}{t'}.
$
Therefore, while the locations of $\mathbf{k_1}, \mathbf{k_2}, \mathbf{k_3}$ and 
$\mathbf{k_4}$ points are independent of $t'$, the last set $\mathbf{k_5}$ exists 
beyond a critical $t'$ value determined by
$
|t'| > \frac{\gamma}{\sin \gamma} \frac{\sin \beta}{2\beta} t 
$
and
$
|t'| > \frac{\gamma}{\sin \gamma} \frac{\sin \alpha}{2\alpha} t,
$
the minimum of which occurs in the ($\alpha, \beta) \to (0,0)$ limit where $|t'| \to t/2$.
The corresponding energy dispersions at the degenerate points are
$\epsilon_{\mathbf{k_1}} = - 2t (\cos \alpha + \cos \beta) - 4t' \cos \gamma$
for the first set,
$\epsilon_{\mathbf{k_2}} = - 2t (\cos \alpha - \cos \beta) + 4t' \cos \gamma$
for the second set,
$\epsilon_{\mathbf{k_3}} =   2t (\cos \alpha - \cos \beta)  + 4t' \cos \gamma$
for the third set,
$\epsilon_{\mathbf{k_4}} =   2t (\cos \alpha + \cos \beta) - 4t' \cos \gamma$
for the fourth set, and
$
\epsilon_{\mathbf{k_5}} = \frac{t^2}{t'} 
\left(
\cos \alpha \frac{\sin \beta}{\beta}
+ \cos \beta \frac{\sin \alpha}{\alpha}
- \frac{\sin \beta}{\beta} \frac{\sin \alpha}{\alpha} \frac{\gamma}{\tan \gamma}
\right) \frac{\gamma}{\sin \gamma}
$
for the last set. 

For instance, in Fig.~\ref{fig:bands}, we show typical band structures as a function 
of $\mathbf{k}$ within the first BZ for (a) $t'/t = 0$, (b) $0.5$, (c) $1$ and (d) $2$, 
when $\alpha = \beta = \pi/4$. Here, since the critical threshold is $t' \approx 0.56 t$, 
the fifth degenerate set clearly emerges only in the latter two figures, where 
$k_5^x = k_5^y$ is approximately given by $0.69\pi$ and $0.59\pi$ when $t'/t$ is $1$ 
and $2$, respectively. When $t'$ is near the threshold value as shown in 
Fig.~\ref{fig:bands}(b), we note that the positive branch becomes almost flat in a 
sizeable $\mathbf{k}$-space region close to the M point along the $\Gamma$ direction.
More importantly, we also note that the bands depend linearly on $\mathbf{k}$ in the
vicinity of all of the touching points, and therefore, they play vital roles in the 
many-particle problem. This is a direct consequence of enhanced single-particle 
density of states by SOC as illustrated next for a number of $(\alpha, \beta)$ values.

\begin{widetext}
\begin{center}
\begin{figure}[htb]
\centerline{
\scalebox{0.345}{\includegraphics{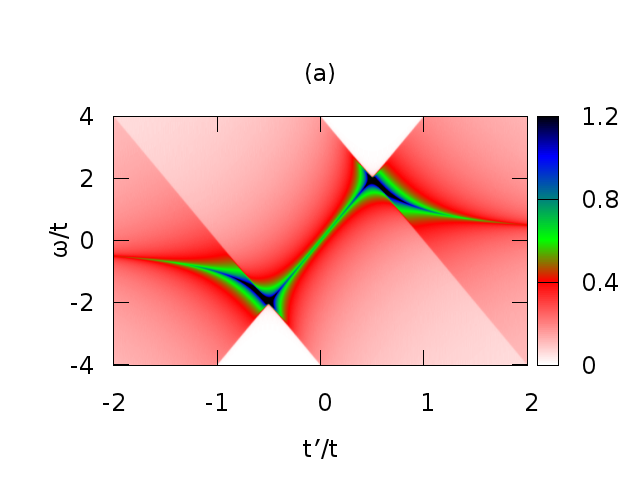}}
\scalebox{0.345}{\includegraphics{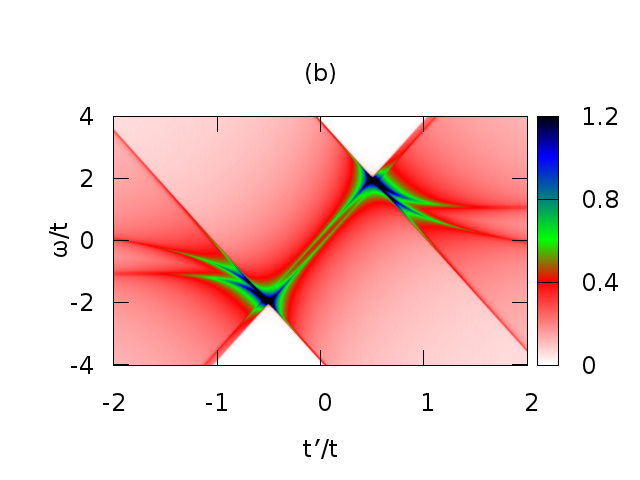}}
}
\vspace{-0.6cm}
\centerline{
\scalebox{0.345}{\includegraphics{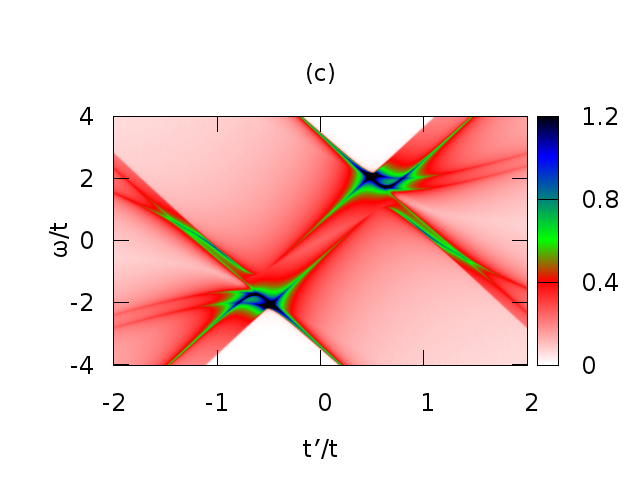}} 
\scalebox{0.345}{\includegraphics{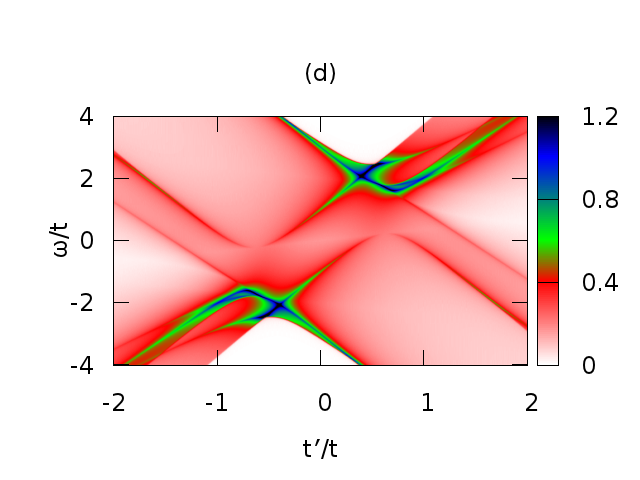}}
}
\vspace{-0.6cm}
\centerline{
\scalebox{0.345}{\includegraphics{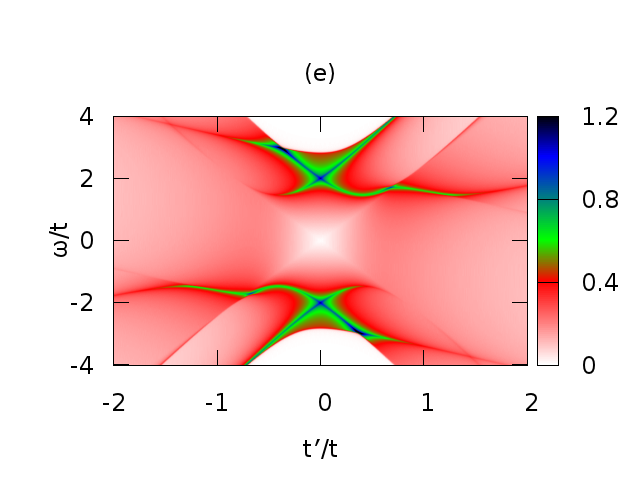}}
\scalebox{0.345}{\includegraphics{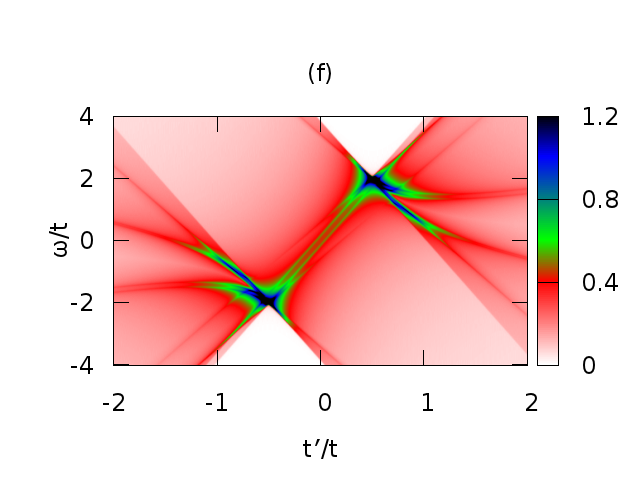}}
}
\vspace{-0.6cm}
\centerline{
\scalebox{0.345}{\includegraphics{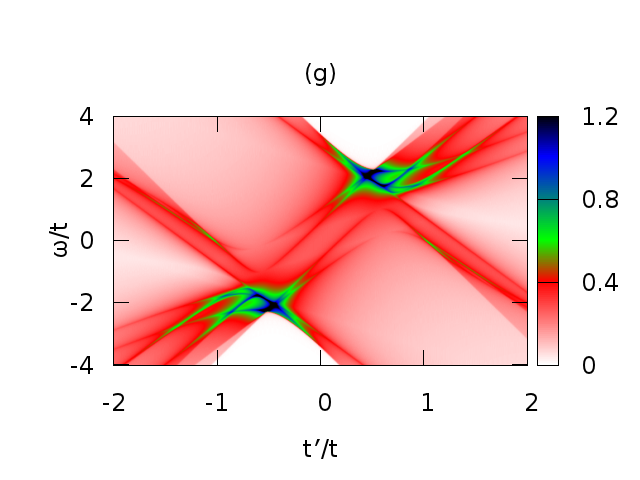}}
\scalebox{0.345}{\includegraphics{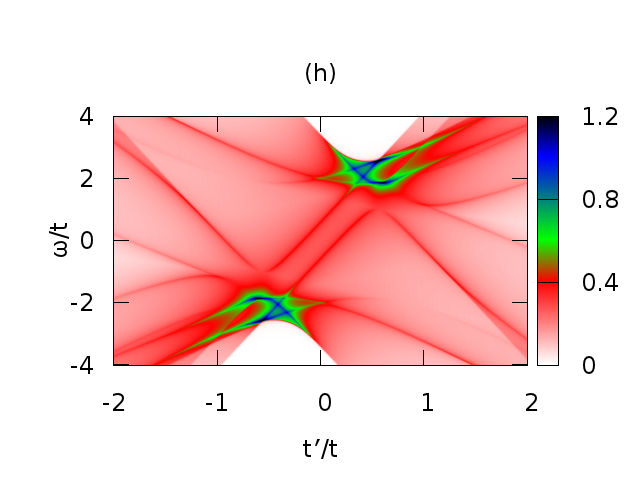}}
}
\caption{\label{fig:dos} (Color online)
The colored maps of the total DoS $D (\omega)$ are shown in units of $1/t$ 
and as functions of energy $\omega/t$ and NNN hopping $t'/t$ for a number 
of $(\alpha, \beta)$ values. 
In the first five figures, we vary $\alpha = \beta$ as (a) $\to 0^+$, (b) $\pi/8$, 
(c) $\pi/4$, (d) $\pi/3$ and (e) $\pi/2$. In the remaining three figures, we set 
$\alpha = \pi/4$ and vary $\beta$ as (f) $\pi/8$, (g) $\pi/3$ and (h) $\pi/2$. 
Note that when $\alpha$ or $\beta$ is strictly zero, the SOC can be 
gauged away, \textit{i.e.}, their $D(\omega)$ are identical to (a).
}
\end{figure}
\end{center}
\end{widetext}
\subsection{Density of States (DoS)}
\label{sec:DoS}

The total DoS $D (\omega) = D_+ (\omega) + D_-(\omega)$ is given by
$
D_\pm (\omega) = (1/M) \sum_\mathbf{k} \delta(\omega - \epsilon_\mathbf{k} \mp |S_\mathbf{k}|),
$
where $\delta(x)$ is the Dirac-delta function, and the following symmetries 
$D (\omega,t') = D(-\omega, -t')$ and $D_-(\omega,t') = D_+(-\omega, -t')$ hold.
We use the broadened definition of the delta function $\pi \delta(x) \to \eta/(x^2 + \eta^2)$
in our numerical calculations with $\eta = 0.01 t$, which is known to be one of the
exact representations of $\delta(x)$ when $\eta \to 0$.

In Fig.~\ref{fig:dos}, we show colored maps of $t D (\omega)$ as functions of 
$\omega/t$ and $t'/t$ for various $(\alpha, \beta)$ values. In the first five figures, 
we study the effects of SOC strength by varying $\alpha = \beta$ 
as (a) $0$, (b) $\pi/8$, (c) $\pi/4$, (d) $\pi/3$ and (e) $\pi/2$.
In the absence of a SOC, shown in Fig.~\ref{fig:dos}(a), the peak position 
of $D(\omega)$ first shifts from $\omega = 0$ to finite values as $|t'|$ increases from 
$0$ to $0.5t$, and then it gradually returns back to $\omega = 0$ in the 
$|t'| \gg t$ limit. In addition, the diverging peak of $D(\omega)$ is considerably broadened 
by finite $t'$ especially around $|t'| \sim t/2$. Therefore, $D(\omega)$ is asymmetric 
around $\omega = 0$ for any given $t' \ne 0$ due to the broken particle/hole symmetry 
by the NNN hopping. On the other hand, in the absence of an NNN hopping, 
while a weak SOC splits the peak of $D(\omega)$ symmetrically around $\omega = 0$ as 
shown in Fig.~\ref{fig:dos}(b), a strong SOC causes two additional peaks located 
at the bottom and top of the spectrum as shown in Figs.~\ref{fig:dos}(c) and~\ref{fig:dos}(d). 
This is a consequence of the broken inversion symmetry by the SOC. 
At the maximally attainable SOC shown in Fig.~\ref{fig:dos}(e), $D(\omega)$ has 
two isolated and broad peaks that are separated by an energy gap at $\omega = 0$.
When both SOC and NNN hopping are coupled, $D(\omega)$ shows 
a peculiar dependence on $\omega$ as shown in Figs.~\ref{fig:dos}(b)-\ref{fig:dos}(e).
For completeness, we show the effects of the symmetry of SOC in the 
remaining three figures, where we set $\alpha = \pi/4$ and vary 
$\beta$ as (f) $\pi/8$, (g) $\pi/3$ and (h) $\pi/2$. Note that when $\alpha$ 
or $\beta$ is strictly zero, the SOC can be gauged away, \textit{i.e.}, their $D(\omega)$ 
are identical to Fig.~\ref{fig:dos}(a).

Having introduced the model Hamiltonian and discussed the effects of SOC both 
on the single-particle problem and on the corresponding DoS, next we briefly go 
over the details of the theoretical formalism for handling the many-body problem.

\subsection{BCS-BEC Evolution}
\label{sec:bcsbec}

In the absence of a gauge field, it has been well-documented in the recent 
literature that the mean-field description of the SF phases given in Eq.~(\ref{eqn:ham})
not only works in the BCS and BEC limits but it also captures much of the 
ground-state properties in the entire BCS-BEC evolution even in two 
dimensions~\cite{kohl11, sommer12, makhalov14, ries15}. 
Motivated by the success of this description in related problems, here we 
apply it to investigate the fate of SF phases under a generic non-Abelian 
gauge field given by Eq.~(\ref{eqn:gauge}).

For the many-body problem, the $\mathbf{k}$-space representation again 
proves to be more convenient, in which case the mean-field Hamiltonian 
can be compactly written as
\begin{align}
H = \frac{1}{2} & \sum_\mathbf{k} \psi_\mathbf{k}^\dagger 
\left( \begin{array}{cccc}
\xi_\mathbf{k}-h & S_\mathbf{k}^\perp & 0 & \Delta \\
S_\mathbf{k}^{\perp *} & \xi_\mathbf{k} + h& -\Delta & 0  \\
0 & -\Delta^* & -\xi_\mathbf{k} + h& S_\mathbf{k}^{\perp *} \\
\Delta^* & 0 &  S_\mathbf{k}^\perp & -\xi_\mathbf{k} - h
\end{array} \right)
\psi_\mathbf{k} \nonumber \\
& + \sum_\mathbf{k} \xi_\mathbf{k} + M \frac{|\Delta|^2}{g},
\label{eqn:ham.k}
\end{align}
where the operator
$
\psi_{\mathbf{k}}^\dagger = [c_{\uparrow \mathbf{k}}^\dagger, 
c_{\downarrow \mathbf{k}}^\dagger,  c_{\uparrow, -\mathbf{k}}, c_{\downarrow, -\mathbf{k}}]
$
denotes the creation and annihilation operators collectively.
The $4 \times 4$ matrix seen in Eq.~(\ref{eqn:ham.k}) is the Hamiltonian matrix, 
say $\mathbf{D}_{\mathbf{k}}$, and it involves the shifted dispersion
$
\xi_\mathbf{k} = \epsilon_\mathbf{k} - \mu,
$
SOC
$
S_\mathbf{k}^\perp = S_\mathbf{k}^x - {\rm i} S_\mathbf{k}^y
$
and complex order parameter $\Delta$ describing the pairing of $\uparrow$ and 
$\downarrow$ atoms in $\mathbf{k}$ space with zero center-of-mass momentum. 
Note that
$
\Delta = g \sum_\mathbf{k} \langle c_{\uparrow \mathbf{k}} c_{\downarrow, -\mathbf{k}} \rangle
$
is uniform in $\mathbf{k}$ space since we allow only for onsite atom-atom 
interactions in real space. The mean-field thermodynamic potential for such 
an $H$ can be written in general as
\begin{equation}
\label{eqn:omega}
\Omega = \frac{T}{2} \sum_{\lambda \mathbf{k}} \ln \left( \frac{1 + X_{\lambda \mathbf{k}}}{2} \right)
+ \sum_\mathbf{k} \xi_\mathbf{k} + M \frac{|\Delta|^2}{g},
\end{equation}
where $T$ is the temperature, $\lambda = \lbrace 1,2,3,4 \rbrace$ labels
the eigenvalues $E_{\lambda \mathbf{k}}$ of the Hamiltonian matrix 
(\textit{i.e.}, the quasiparticle/quasihole excitation energies) and
$
X_{\lambda \mathbf{k}} = \tanh[E_{\lambda \mathbf{k}}/(2T)].
$ 
The Boltzmann constant $k_B$ is set to unity throughout this paper. 
The eigenvalues of our $\mathbf{D}_{\mathbf{k}}$ have relatively simple 
analytic forms, and can be compactly written as
\begin{equation}
\label{eqn:Ek}
E_{\lambda \mathbf{k}} = s_\lambda \sqrt{\xi_\mathbf{k}^2+h^2+|\Delta|^2+|S_\mathbf{k}^\perp|^2 + 2 p_\lambda Z_\mathbf{k}},
\end{equation}
where $s_{1,3} = p_{3,4} = +1$ denote the quasiparticle and $s_{2,4} = p_{1,2} = -1$ 
denote the quasihole branches which are particle/hole symmetric around zero energy, and 
$
Z_\mathbf{k} = \sqrt{(\xi_\mathbf{k}^2 + |\Delta|^2)h^2 + |S_\mathbf{k}^\perp|^2 \xi_\mathbf{k}^2}.
$

Given $\Omega$, the self-consistency equations are simply obtained by imposing 
the following conditions: $\partial \Omega / \partial |\Delta| = 0$ gives an expression 
for the order parameter,  $\partial \Omega / \partial \mu = -N$ gives the total number 
of $\uparrow$ and $\downarrow$ atoms, and $\partial \Omega / \partial h = - P N$ 
determines their polarization. This procedure leads to
\begin{align}
\label{eqn:gap}
\frac{2M |\Delta|}{g} &= \frac{1}{4} \sum_{\lambda \mathbf{k}} \frac{\partial E_{\lambda \mathbf{k}}}{\partial |\Delta|} \left( X_{\lambda \mathbf{k}} - 1 \right), \\
\label{eqn:ntot}
N_\uparrow + N_\downarrow &= \frac{1}{4} \sum_{\lambda \mathbf{k}} \left[1 + \frac{\partial E_{\lambda \mathbf{k}}}{\partial \mu} \left( X_{\lambda \mathbf{k}} -1 \right) \right], \\
\label{eqn:ndif}
N_\uparrow - N_\downarrow &= \frac{1}{4} \sum_{\lambda \mathbf{k}} \frac{\partial E_{\lambda \mathbf{k}}}{\partial h} \left( X_{\lambda \mathbf{k}} -1 \right),
\end{align}
where the derivatives are
$
\partial E_{\lambda \mathbf{k}} / \partial |\Delta| = (1 + p_\lambda h^2/Z_\mathbf{k} ) |\Delta| / E_{\lambda \mathbf{k}}
$
for the order parameter,
$
\partial E_{\lambda \mathbf{k}} / \partial \mu = - [1 + p_\lambda ( h^2+|S_\mathbf{k}^\perp|^2 )/Z_\mathbf{k} ] \xi_\mathbf{k} / E_{\lambda \mathbf{k}}
$
for the chemical potential, and
$
\partial E_{\lambda \mathbf{k}} / \partial h = - [1 + p_\lambda ( \xi_\mathbf{k}^2+|\Delta|^2 )/Z_\mathbf{k} ] h / E_{\lambda \mathbf{k}}
$
for the Zeeman field. Since $|\Delta|$, $\mu$ and $h$ are coupled for a given 
$t', g, \alpha$, $\beta$ and $T$, Eqs.~(\ref{eqn:gap}), (\ref{eqn:ntot}) and (\ref{eqn:ndif})
need to be solved simultaneously for a self-consistent solution. 
Except for some limiting cases, the best way to achieve such solutions is 
through numerical means, \textit{e.g.}, via an iterative approach as discussed 
below in Sec.~\ref{sec:numerics}.

\section{Topological Phase Transitions}
\label{sec:topo} 

Since it is possible to classify and distinguish all of the SF phases by the 
$\mathbf{k}$-space topology of their zero-energy quasiparticle/quasihole 
excitations~\cite{iskin13a}, next we analyze $E_{\lambda \mathbf{k}}$ for 
gapped/gapless solutions and gain some analytical insight before plunging 
into the numerical results.

\subsection{Gapless Superfluids}
\label{sec:gapless} 

Equation~(\ref{eqn:Ek}) clearly shows that while $E_{3 \mathbf{k}}$ and 
$E_{4 \mathbf{k}}$ never vanish and are always gapped as long as $|\Delta| \ne 0$, 
$E_{1 \mathbf{k}}$ and $E_{2 \mathbf{k}}$ may vanish and can become gapless 
at some special $\mathbf{k_0}$ points in $\mathbf{k}$ space. After a little bit of algebra, 
it can be easily verified that the condition $E_{1(2) \mathbf{k}} = 0$ boils down to 
two conditions: $|S_\mathbf{k}^\perp| = 0$ together with 
$(\xi_\mathbf{k} + h) (\xi_\mathbf{k} - h) + |\Delta|^2 = 0$.
Therefore, while the number and locations of $\mathbf{k_0}$ depend only 
on $t', \alpha$ and $\beta$, the critical Zeeman fields required
$
h_{\mathbf{k}_0} = \sqrt{\xi_\mathbf{k_0}^2 + |\Delta|^2}
$ 
depend on $t', \alpha, \beta, \epsilon_\mathbf{k_0}, \mu$ and $g$.
Note that the presence of a SOC is not one of the essential ingredients that 
give rise to quantum phase transitions between SF phases with distinct 
$\mathbf{k}$-space topologies, and that these transitions are solely driven by 
$h \ge h_{\mathbf{k}_0}$ causing eventually $P \ne 0$ even in its absence.

For instance, we may have up to five sets of $\mathbf{k_0}$ points satisfying 
the first gapless condition $|S_\mathbf{k_0}^\perp| = 0$, which is discussed at
length in Sec.~\ref{sec:bands}. Based on this analysis, it seems possible 
to have quantum phase transitions between SF phases with distinct 
$\mathbf{k}$-space topologies at five different $h_\mathbf{k_0}$ values.
However, these distinct gapless-SF phases may only be realized 
assuming $|\Delta| \ne 0$ still pertains at sufficiently high Zeeman fields 
satisfying simultaneously the second gapless condition $h = h_{\mathbf{k}_0}$.
The verification of this condition has to wait until Sec.~\ref{sec:numerics}, 
where we present our self-consistent solutions.

\subsection{Topological Superfluids}
\label{sec:chern} 

In the presence of a SOC, in addition to the gapped/gapless nature of 
$E_{\lambda \mathbf{k}}$ and the $\mathbf{k}$-space topology of the zero-energy 
quasiparticle/quasihole excitations, SF phases may further be characterized 
as topological or not depending on the integer value of the topological 
invariant of the system. By definition, a topologically-invariant quantity may only 
change due to a change in the topology of the system, \textit{e.g.}, by opening 
or closing of an energy gap in $\mathbf{k}$ space, and therefore, the lowest-energy 
excitations of the system play a crucial role. To keep track of the possible 
changes in the topological invariant across the critical field $h_\mathbf{k_0}$, 
here we follow Bellissard's proposal~\cite{bellissard95} offering a simple 
and transparent way of calculating the changes in the CN.

For this purpose, we reduce the $4 \times 4$ matrix $\mathbf{D}_\mathbf{k}$ to 
an effective $2 \times 2$ matrix $\mathbf{D}_{\mathbf{k} \mathbf{k_0}}^{\rm eff}$,
describing the Hamiltonian dynamics in the neighborhood of 
$\mathbf{k} = \mathbf{k_0}$ and $h = h_\mathbf{k_0}$.
First, setting $\mathbf{k} = \mathbf{k_0}$ and $S_\mathbf{k_0}^\perp = 0$ in 
$\mathbf{D}_\mathbf{k}$, we obtain much simpler expressions
$
E_{1(2) \mathbf{k_0}} = \pm \left( h_\mathbf{k_0} - h \right)
$
for the $\lambda = \{1,2\}$ and
$
E_{3(4) \mathbf{k_0}} = \pm \left(h_\mathbf{k_0} + h \right)
$
for the $\lambda = \{3,4\}$ branches of the quasiparticle/quasihole symmetric excitation 
spectrum. Then, the corresponding eigenvectors $\mathbf{f}_{\lambda \mathbf{k_0}}$ 
for the zero-energy eigenvalues can be written as
$
\mathbf{f}_{1 \mathbf{k_0}}^{\rm T} 
= C_{1 \mathbf{k_0}} \left( \frac{h_\mathbf{k_0} + \xi_\mathbf{k_0}}{\Delta^*}, 0, 0, 1 \right)^{\rm T}
$
for $E_{1 \mathbf{k_0}} = 0$ and
$
\mathbf{f}_{2 \mathbf{k_0}}^{\rm T} 
= C_{2 \mathbf{k_0}} \left( 0, \frac{h_\mathbf{k_0} - \xi_\mathbf{k_0}}{\Delta^*}, 1, 0 \right)^{\rm T}
$
for $E_{2 \mathbf{k_0}} = 0$. Here, ${\rm T}$ is the transposition operator, and 
$
C_{1(2) \mathbf{k_0}} = |\Delta|/\sqrt{2h_\mathbf{k_0} (h_\mathbf{k_0} \pm \xi_\mathbf{k_0})}
$
are the normalization factors.
Finally, projecting $\mathbf{D}_\mathbf{k}$ onto the two-dimensional 
zero-energy subspace of $\mathbf{f}_{1 \mathbf{k_0}}$ and $\mathbf{f}_{2 \mathbf{k_0}}$ 
eigenvectors gives the approximate Hamiltonian matrix $\mathbf{D}_{\mathbf{k} \mathbf{k_0}}^{\rm eff}$ 
near the zeros, whose elements are
$
D_{\mathbf{k} \mathbf{k_0}}^{ij} = \mathbf{f}_{i \mathbf{k_0}}^\dagger \mathbf{D}_\mathbf{k} \mathbf{f}_{j \mathbf{k_0}}.
$ 
Therefore, $\mathbf{D}_{\mathbf{k} \mathbf{k_0}}^{\rm eff}$ can be compactly written as
$
\mathbf{D}_{\mathbf{k}\mathbf{k_0}}^{\rm eff} 
= d_0 \sigma_0 + \mathbf{d}_{\mathbf{k}\mathbf{k_0}} \cdot \vec{\sigma},
$
where $d_0 = 0$ due to quasiparticle/quasihole symmetry, and the elements of the vector
$
\mathbf{d}_{\mathbf{k}\mathbf{k_0}} = (d_{\mathbf{k}\mathbf{k_0}}^x, 
d_{\mathbf{k}\mathbf{k_0}}^y, d_{\mathbf{k}\mathbf{k_0}}^z)
$
are
\begin{align}
\label{eqn:dkx}
d_{\mathbf{k}\mathbf{k_0}}^x &= \frac{|\Delta|}{h_\mathbf{k_0}} S_\mathbf{k}^x, \\
\label{eqn:dky}
d_{\mathbf{k}\mathbf{k_0}}^y &= \frac{|\Delta|}{h_\mathbf{k_0}} S_\mathbf{k}^y, \\
\label{eqn:dkz}
d_{\mathbf{k}\mathbf{k_0}}^z &= \frac{|\Delta|^2}{h_\mathbf{k_0}} \left(1 - \frac{h h_\mathbf{k_0} - \xi_\mathbf{k} \xi_\mathbf{k_0}}{|\Delta|^2} \right).
\end{align}
These expressions satisfy~\cite{bellissard95} $d_0 = (E_{1 \mathbf{k}} + E_{2 \mathbf{k}})/2$ and
$
|\mathbf{d}_{\mathbf{k}\mathbf{k_0}}| \approx (E_{1 \mathbf{k}} - E_{2 \mathbf{k}})/2
$ 
in such a way that $\mathbf{d}_{\mathbf{k}\mathbf{k_0}}$ vanishes precisely 
at the locations of zeros.

According to Bellissard's proposal~\cite{bellissard95}, the change in the CN at 
$h = h_\mathbf{k_0}$ is given by a sum over the Berry indices at all touching points
\begin{align}
\label{eqn:CN}
\Delta {\rm CN} (h_\mathbf{k_0}) &= \sum_\mathbf{k_0} {\rm sign} \{ \det [\mathbf{J}_\mathbf{f}(\mathbf{k_0}) ] \}, \\
\label{eqn:detJ}
\det [\mathbf{J}_\mathbf{f} (\mathbf{k_0})] &= \left[ 
\frac{\partial d_{\mathbf{k}\mathbf{k_0}}^x}{\partial k_x} \frac{\partial d_{\mathbf{k}\mathbf{k_0}}^y}{\partial k_y}
- \frac{\partial d_{\mathbf{k}\mathbf{k_0}}^y}{\partial k_x} \frac{\partial d_{\mathbf{k}\mathbf{k_0}}^x}{\partial k_y} 
\right]_\mathbf{k_0}.
\end{align}
From the topological perspective, the unit vector $\widehat{\mathbf{d}}_{\mathbf{k}\mathbf{k_0}}$ 
maps $\mathbf{k}$ space ($T^2$) to an infinitesimally-small unit sphere ($S^2$) enclosing 
the zero at $\mathbf{k_0}$, and $\mathbf{J}_\mathbf{f} (\mathbf{k_0})$ 
is the Jacobian matrix for the corresponding transformation. The CN is physically
the number of times that $\widehat{\mathbf{d}}_{\mathbf{k}\mathbf{k_0}}$ winds 
around $S^2$ as $\mathbf{k}$ varies, and therefore, it is also called the winding number.
Note that, since $\partial d_{\mathbf{k}\mathbf{k_0}}^x / \partial h = 0$,  
$\partial d_{\mathbf{k}\mathbf{k_0}}^y / \partial h = 0$ and 
$\partial d_{\mathbf{k} \mathbf{k_0}}^z / \partial h = - 1$ in our problem, 
the Jacobian of the transformation satisfies
$
\det [\mathbf{J}_\mathbf{f} (h_0, \mathbf{k_0}) ] = - \det [\mathbf{J}_\mathbf{f} (\mathbf{k_0}) ].
$

For our square lattice, using Eqs.~(\ref{eqn:skx.s}) and~(\ref{eqn:sky.s}) 
in Eq.~(\ref{eqn:detJ}), we find 
\begin{align}
\det [\mathbf{J}_\mathbf{f} (\mathbf{k_1})]  &= \frac{4|\Delta|^2 t^2}{h_\mathbf{k_1}^2} \left[
\sin \alpha \sin \beta 
+ 4 \alpha \beta \frac{\sin^2 \gamma} {\gamma^2} \frac{t'^2}{t^2} \right. \nonumber \\
&+ \left. 2 (\alpha \sin \beta + \beta \sin \alpha) \frac{\sin \gamma} {\gamma} \frac{t'}{t}
\right]
\label{eqn:j1.s}
\end{align}
for the first set of zeros. Simply substitute $\sin \alpha \to -\sin \alpha$
in this expression to obtain $\det [\mathbf{J}_\mathbf{f} (\mathbf{k_2})]$,
substitute $\sin \beta \to -\sin \beta$ to obtain $\det [\mathbf{J}_\mathbf{f} (\mathbf{k_3})]$,
and substitute $\sin \alpha \to -\sin \alpha$ together with $\sin \beta \to -\sin \beta$ 
to obtain $\det [\mathbf{J}_\mathbf{f} (\mathbf{k_4})]$.
For the last set of zeros, we likewise obtain
\begin{align}
\det [\mathbf{J}_\mathbf{f} (\mathbf{k_5})]  = - \frac{16|\Delta|^2 t'^2}{h_\mathbf{k_5}^2} & \alpha \beta
\left( \frac{\sin^2 \gamma} {\gamma^2} - \frac{\sin^2 \beta} {4 \beta^2} \frac{t^2}{t'^2}  \right) \nonumber \\
\times & \left( \frac{\sin^2 \gamma} {\gamma^2} - \frac{\sin^2 \alpha} {4 \alpha^2} \frac{t^2}{t'^2} \right),
\label{eqn:j5.s}
\end{align}
which is always negative when $\mathbf{k_5}$ is relevant. 
Therefore, $\Delta {\rm CN} (h_\mathbf{k_5}) = -4$ for all parameters as long as 
$|t'|/t$ is beyond the critical threshold value given in Sec.~\ref{sec:bands}.
Since the total change in the CN must add up to zero in the normal phase when $h \gg t$, 
this suggests that $\Delta {\rm CN} (h_\mathbf{k_1}) = \Delta {\rm CN} (h_\mathbf{k_2})
= \Delta {\rm CN} (h_\mathbf{k_3}) = \Delta {\rm CN} (h_\mathbf{k_4}) = + 1$
for all parameters, when $\mathbf{k_5}$ is relevant. 
This is best revealed in the following limits.

\textit{(i)} When $t' = 0$, Eq.~(\ref{eqn:j1.s}) reduces to
$
\det [\mathbf{J}_\mathbf{f} (\mathbf{k_1})] = - \det [\mathbf{J}_\mathbf{f} (\mathbf{k_2})] 
= - \det [\mathbf{J}_\mathbf{f} (\mathbf{k_3})] = \det [\mathbf{J}_\mathbf{f} (\mathbf{k_4})] 
=  4|\Delta|^2 t^2 \sin \alpha \sin \beta / h_\mathbf{k_{1,2,3,4}}^2,
$
for which the corresponding energy dispersions are given by
$\epsilon_\mathbf{k_1} = - \epsilon_\mathbf{k_4} = -2t(\cos \alpha + \cos \beta)$
and
$\epsilon_\mathbf{k_2} = - \epsilon_\mathbf{k_3} = -2t(\cos \alpha - \cos \beta)$.
Note that $\mathbf{k_5}$ is irrelevant in this limit. 
Therefore, $\Delta {\rm CN} (h_\mathbf{k_1}) = \Delta {\rm CN} (h_\mathbf{k_4}) = + 1$ 
and $\Delta {\rm CN} (h_\mathbf{k_2}) = \Delta {\rm CN} (h_\mathbf{k_3}) = -1$.
However, $h_\mathbf{k_0}$ are four-fold (two-fold) degenerate when 
$\alpha = \pi/2$ and (or) $\beta = \pi/2$ with opposite contributions to the 
change in CN, and therefore, the total change in the CN vanishes for any $\mu$.

\textit{(ii)} When $t = 0$, Eqs.~(\ref{eqn:j1.s}) and~(\ref{eqn:j5.s}) reduce to
$
\det [\mathbf{J}_\mathbf{f} (\mathbf{k_1})] = \det [\mathbf{J}_\mathbf{f} (\mathbf{k_2})] 
= \det [\mathbf{J}_\mathbf{f} (\mathbf{k_3})] = \det [\mathbf{J}_\mathbf{f} (\mathbf{k_4})] 
= - \det [\mathbf{J}_\mathbf{f} (\mathbf{k_5})] 
=  16|\Delta|^2 \alpha \beta \sin^2 \gamma t'^2 / (\gamma^2 h_\mathbf{k_{1,2,3,4,5}}^2),
$
for which the corresponding energy dispersions are given by 
$\epsilon_\mathbf{k_1} = - \epsilon_\mathbf{k_2} = -\epsilon_\mathbf{k_3} 
=  \epsilon_\mathbf{k_4} = -4t' \cos \gamma$
and $\epsilon_\mathbf{k_5} = 0$.
When $\mu \ne 0$, the fields $h_\mathbf{k_1} = h_\mathbf{k_4}$ and $h_\mathbf{k_2} = h_\mathbf{k_3}$
are two-fold degenerate, and therefore,
$\Delta {\rm CN} (h_\mathbf{k_{1,4}}) = \Delta {\rm CN} (h_\mathbf{k_{2,3}}) = +2$ 
and $\Delta {\rm CN}(h_\mathbf{k_5}) = -4$.
At $\mu = 0$, however, $h_\mathbf{k_0}$ are four-fold degenerate, and 
$\Delta {\rm CN}$ changes from $0$ to $-4$ to $0$ as $h$ increases from 
$0$ to $h_\mathbf{k_5} = |\Delta|$ to $h_\mathbf{k_{1,2,3,4}} = \sqrt{16t'^2 \cos^2 \gamma + |\Delta|^2}$.

\textit{(iii)} When $\alpha =  \beta$, Eq.~(\ref{eqn:j1.s}) reduces to
$
\det [\mathbf{J}_\mathbf{f} (\mathbf{k_1})] = 4|\Delta|^2 [t \sin \alpha + \sqrt{2} t' \sin(\sqrt{2} \alpha)]^2 / h_\mathbf{k_1}^2,
$
and we find 
$
\det [\mathbf{J}_\mathbf{f} (\mathbf{k_2})] = \det [\mathbf{J}_\mathbf{f} (\mathbf{k_3})] 
= - 4|\Delta|^2 [t^2 \sin^2 \alpha - 2 t'^2 \sin^2 (\sqrt{2} \alpha)] / h_\mathbf{k_{2,3}}^2
$
for the second and third sets of zeros, and
$
\det [\mathbf{J}_\mathbf{f} (\mathbf{k_4})] = 4|\Delta|^2 [t \sin \alpha - \sqrt{2} t' \sin(\sqrt{2} \alpha)]^2 / h_\mathbf{k_4}^2
$
for the fourth set of zeros. In addition, Eq.~(\ref{eqn:j5.s}) reduces to
$
\det [\mathbf{J}_\mathbf{f} (\mathbf{k_5})] = - 8|\Delta|^2 
\left[\frac{\sin^2 \alpha}{\sin (\sqrt{2} \alpha)} \frac{t^2}{2t'} - \sin (\sqrt{2} \alpha) t' \right]^2 / h_\mathbf{k_5}^2
$
which is relevant for $|t'| > t \sin \alpha / [\sqrt{2} \sin (\sqrt{2} \alpha)]$.
The corresponding energy dispersions are given by 
$\epsilon_\mathbf{k_1} = -4t \cos \alpha - 4t' \cos (\sqrt{2} \alpha)$,
$\epsilon_\mathbf{k_2} = \epsilon_\mathbf{k_3} = 4t' \cos (\sqrt{2} \alpha)$,
$\epsilon_\mathbf{k_4} = 4t \cos \alpha - 4t' \cos (\sqrt{2} \alpha)$ and
$\epsilon_\mathbf{k_5} = \sqrt{2} \frac{t^2}{t'}
\left[ \frac{\sin (2\alpha)}{\sin (\sqrt{2}\alpha)} - \frac{\sqrt{2} \sin^2 \alpha}{\sin (\sqrt{2} \alpha) \tan (\sqrt{2} \alpha)} \right]$.
Therefore, since $h_\mathbf{k_2} = h_\mathbf{k_3}$ is two-fold degenerate, 
we find $\Delta {\rm CN}(h_\mathbf{k_1}) = \Delta {\rm CN}(h_\mathbf{k_4}) = +1$,
$\Delta {\rm CN}(h_\mathbf{k_{2,3}}) = \pm 2$ and $\Delta {\rm CN}(h_\mathbf{k_5}) = -4$,
depending on whether $\mathbf{k_5}$ is relevant or not.

\begin{widetext}
\begin{center}

\begin{figure}[htb]
\vspace{-1cm}
\centerline{
\scalebox{0.345}{\includegraphics{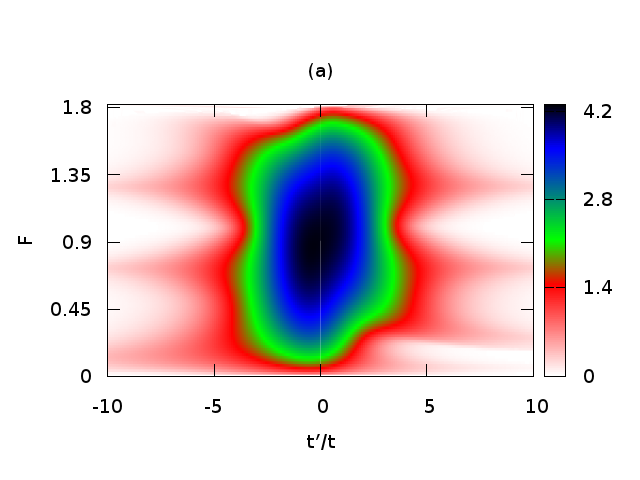}}
\scalebox{0.3}{\includegraphics{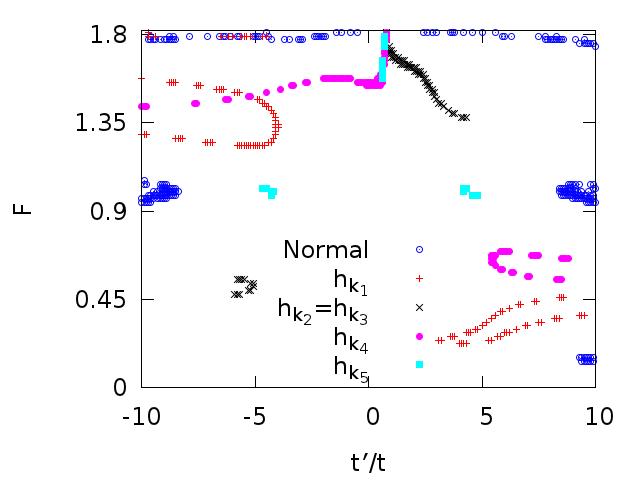}} 
}
\vspace{-0.23cm}
\centerline{
\scalebox{0.345}{\includegraphics{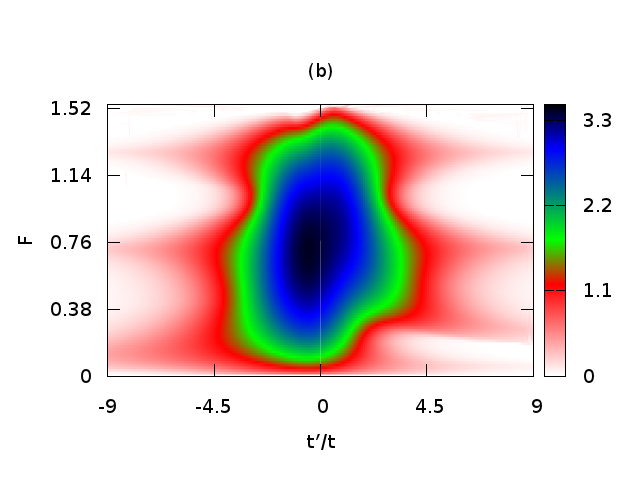}}
\scalebox{0.3}{\includegraphics{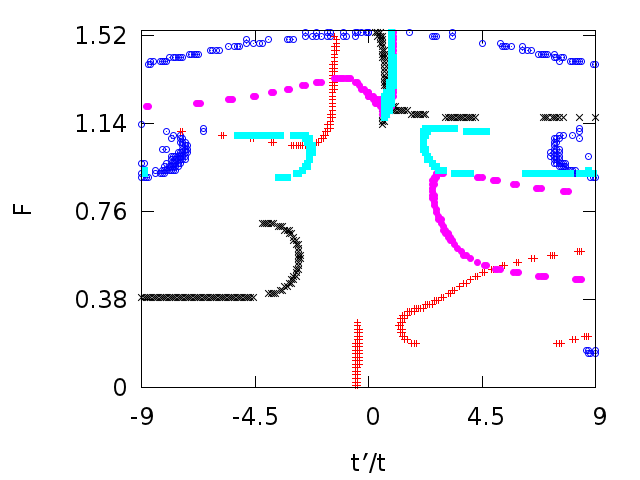}} 
}
\vspace{-0.23cm}
\centerline{
\scalebox{0.345}{\includegraphics{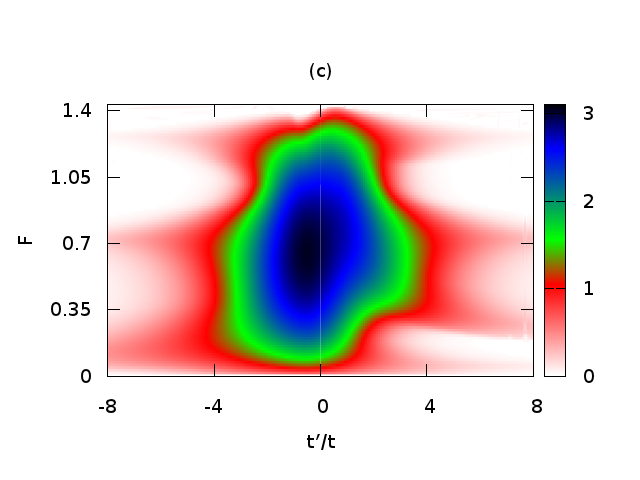}}
\scalebox{0.3}{\includegraphics{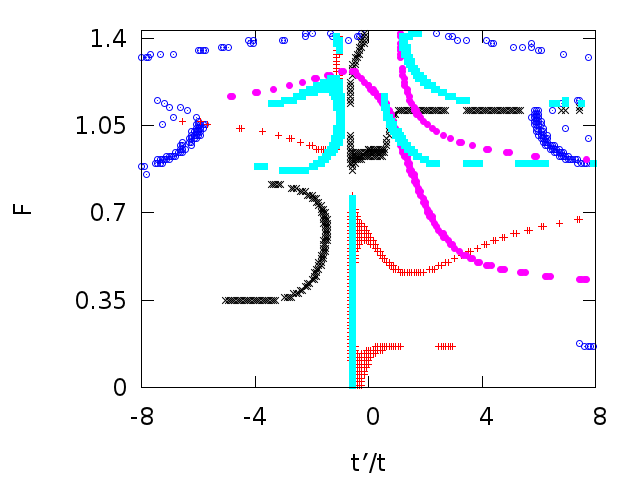}} 
}
\vspace{-0.23cm}
\centerline{
\scalebox{0.345}{\includegraphics{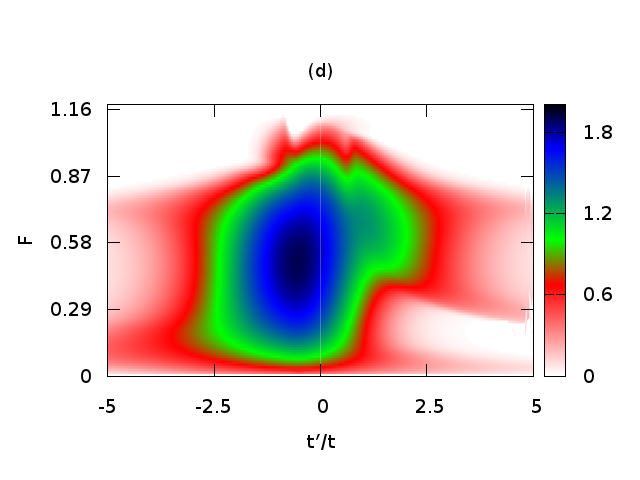}}
\scalebox{0.3}{\includegraphics{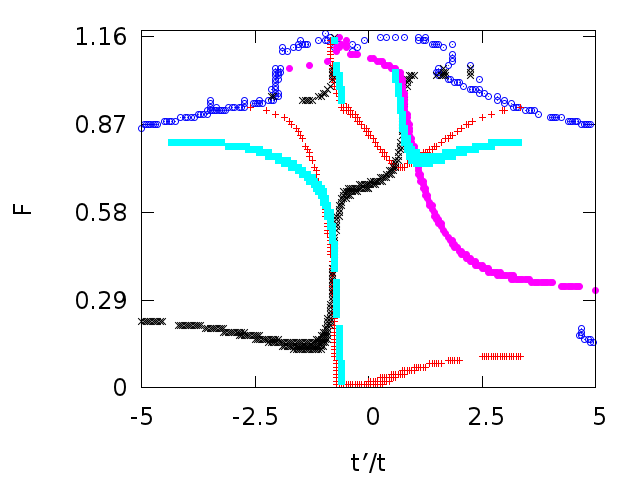}} 
}
\caption{\label{fig:P} (Color online)
The colored maps of the SF order parameter $|\Delta|/t$ are shown 
as functions of total particle filling $F$ and NNN hopping $t'/t$ for 
(a) $P = 0.1$, (b) $0.3$, (c) $0.4$ and (d) $0.7$,
where $\alpha = \beta = \pi/4$ and $g = 10t$.
The corresponding boundaries for the topological quantum phase transitions 
are determined by the condition 
$
h = h_{\mathbf{k}_i} = \sqrt{\xi_\mathbf{k_i}^2 + |\Delta|^2},
$
and they are also extracted and shown on the right panel.
}
\end{figure}

\begin{figure}[htb]
\vspace{-1cm}
\centerline{
\scalebox{0.345}{\includegraphics{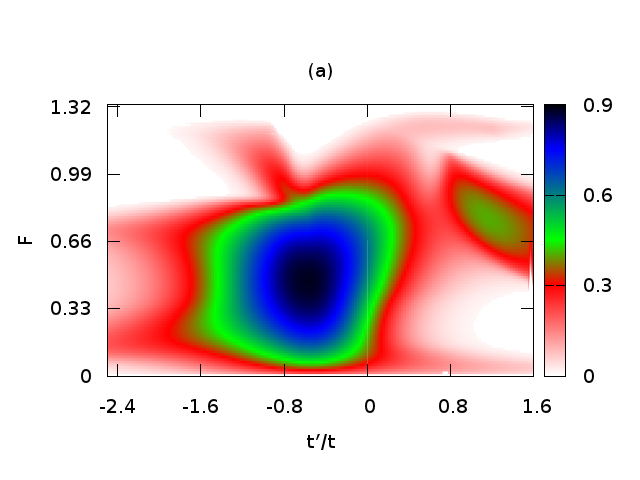}}
\scalebox{0.3}{\includegraphics{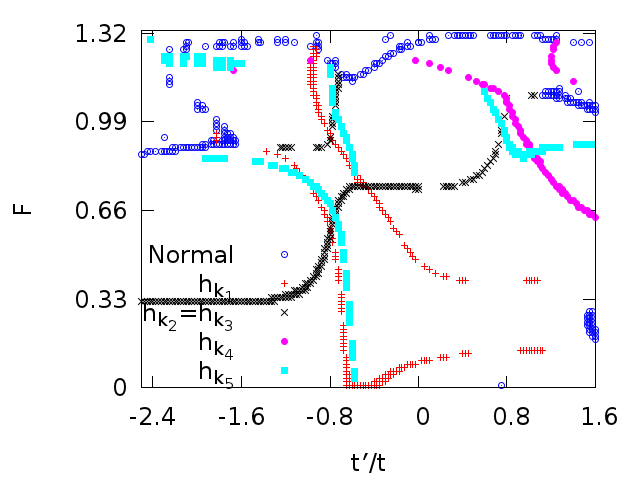}} 
}
\vspace{-0.23cm}
\centerline{
\scalebox{0.345}{\includegraphics{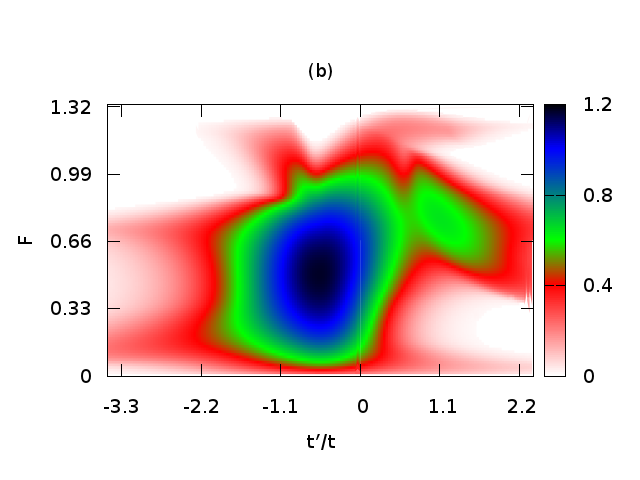}}
\scalebox{0.3}{\includegraphics{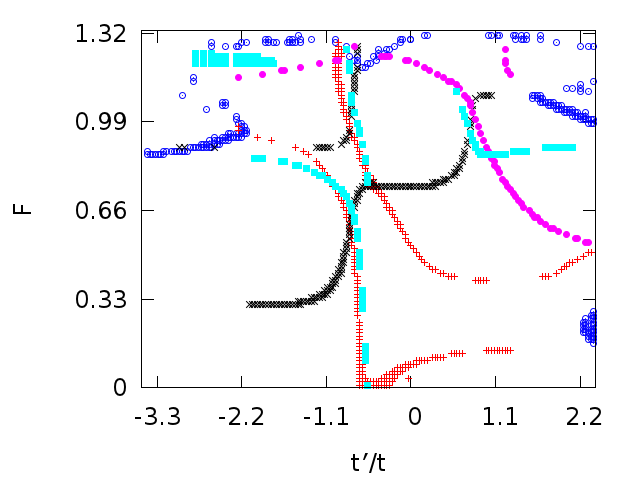}} 
}
\vspace{-0.23cm}
\centerline{
\scalebox{0.345}{\includegraphics{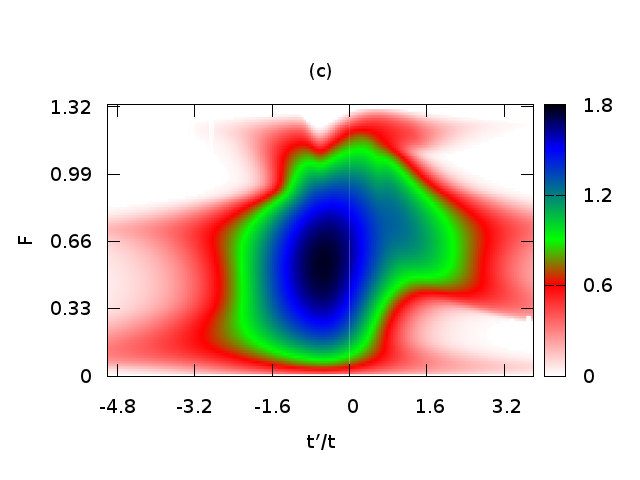}}
\scalebox{0.3}{\includegraphics{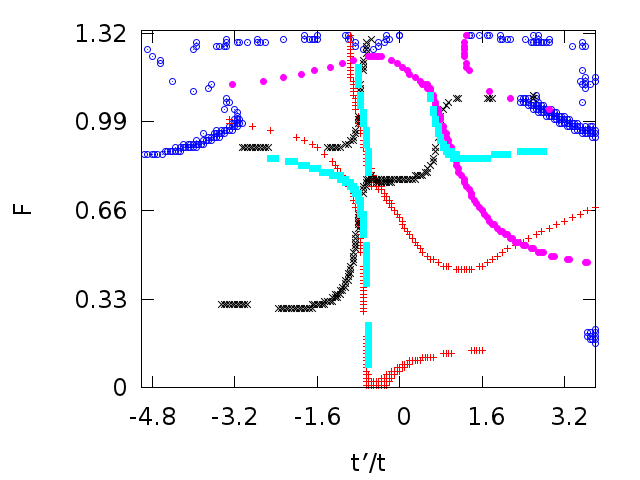}} 
}
\vspace{-0.23cm}
\centerline{
\scalebox{0.345}{\includegraphics{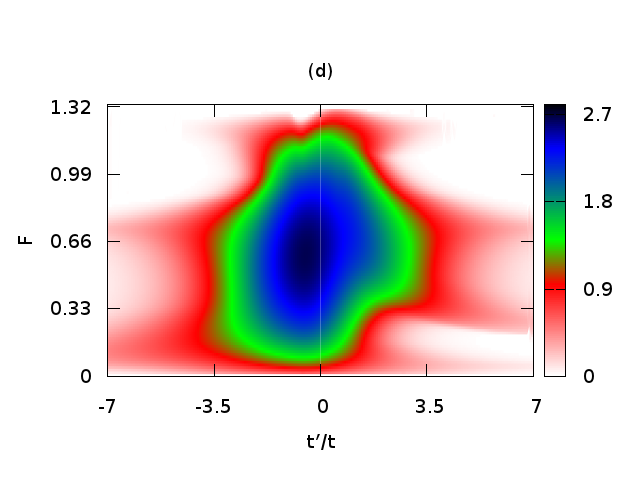}}
\scalebox{0.3}{\includegraphics{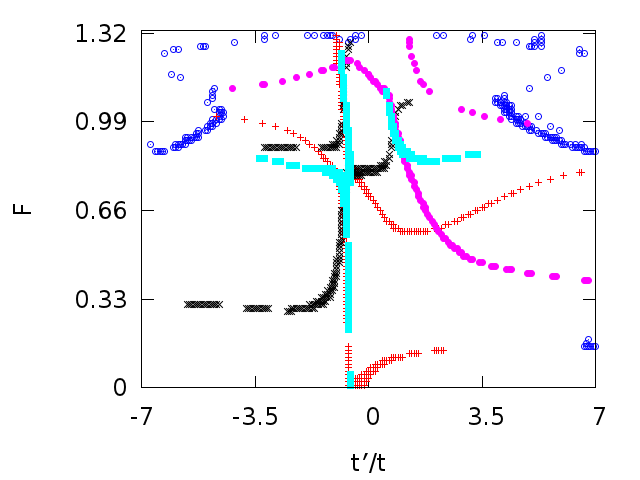}} 
}
\caption{\label{fig:g} (Color online)
The colored maps of the SF order parameter $|\Delta|/t$ are shown 
as functions of total particle filling $F$ and NNN hopping $t'/t$ for 
(a) $g = 4t$, (b) $5t$, (c) $7t$  and (d) $10t$,
where $\alpha = \beta = \pi/4$ and $P = 0.5$.
The corresponding boundaries for the topological quantum phase transitions 
are determined by the condition 
$
h = h_{\mathbf{k}_i} = \sqrt{\xi_\mathbf{k_i}^2 + |\Delta|^2},
$
and they are also extracted and shown on the right panel.
}
\end{figure}

\begin{figure}[htb]
\vspace{-1cm}
\centerline{
\scalebox{0.345}{\includegraphics{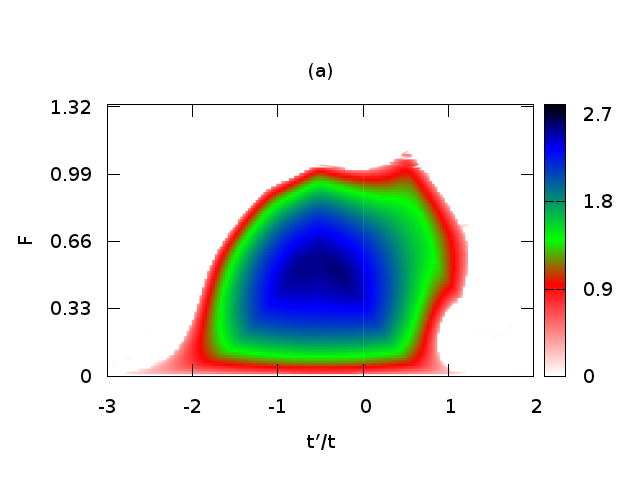}}
\scalebox{0.3}{\includegraphics{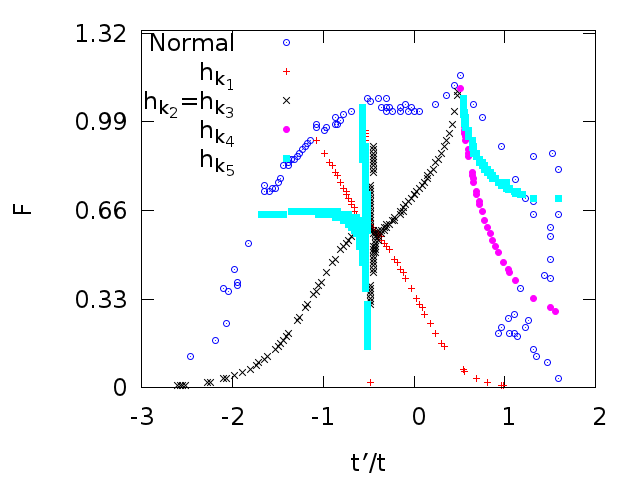}} 
}
\vspace{-0.23cm}
\centerline{
\scalebox{0.345}{\includegraphics{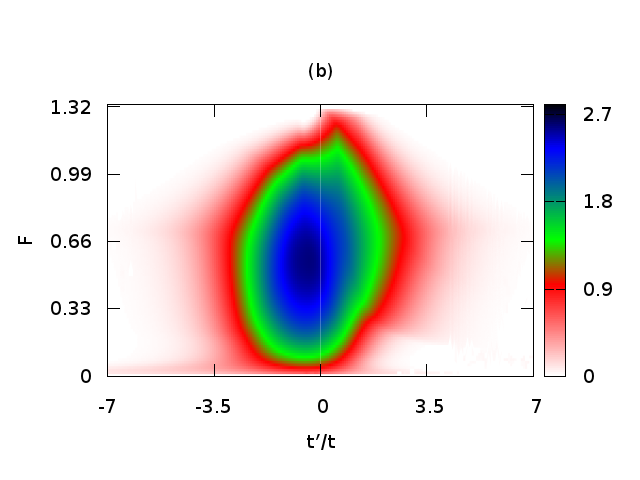}}
\scalebox{0.3}{\includegraphics{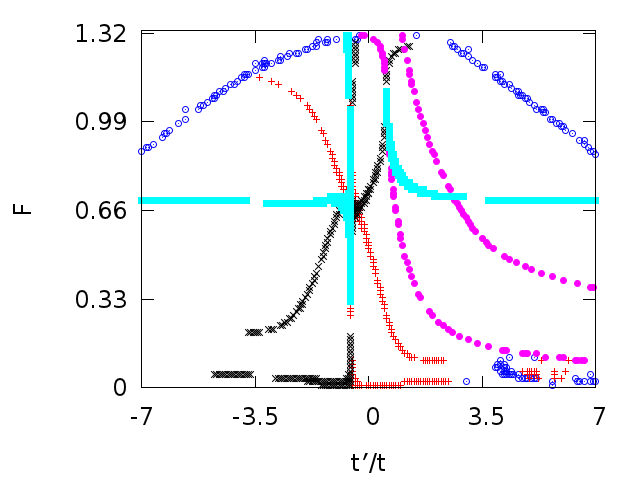}} 
}
\vspace{-0.23cm}
\centerline{
\scalebox{0.345}{\includegraphics{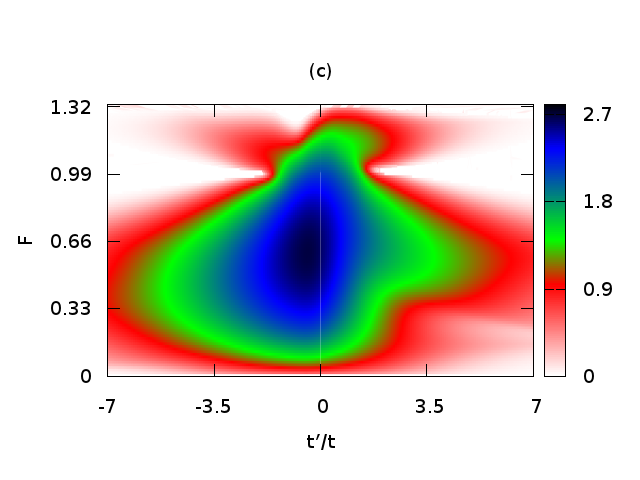}}
\scalebox{0.3}{\includegraphics{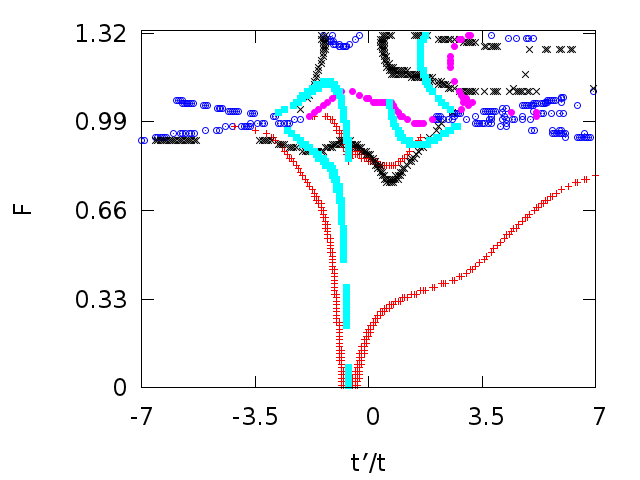}} 
}
\vspace{-0.23cm}
\centerline{
\scalebox{0.345}{\includegraphics{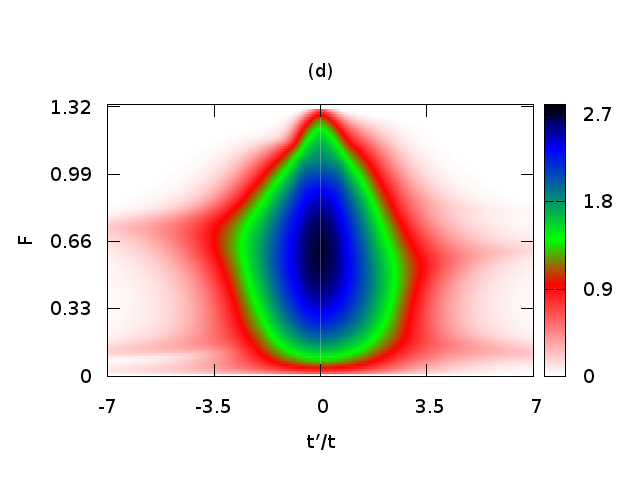}}
\scalebox{0.3}{\includegraphics{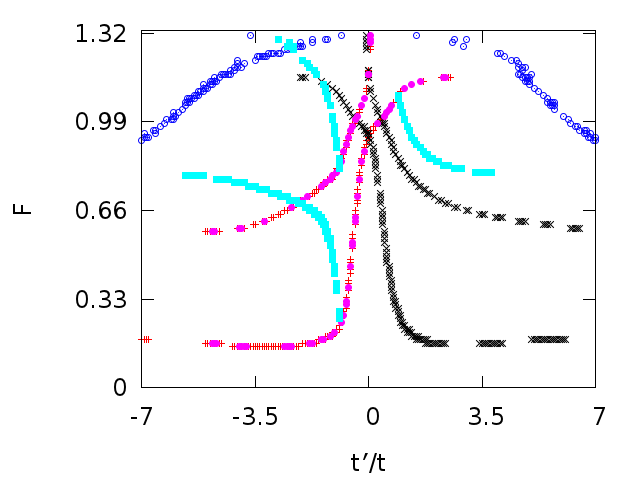}} 
}
\caption{\label{fig:ab} (Color online)
The colored maps of the SF order parameter $|\Delta|/t$ are shown 
as functions of total particle filling $F$ and NNN hopping $t'/t$ for 
(a) $\alpha = \beta \to 0^+$, (b) $\pi/8$, (c) $\pi/3$  and (d) $\pi/2$,
where $g = 10t$ and $P = 0.5$.
The corresponding boundaries for the topological quantum phase transitions 
are determined by the condition 
$
h = h_{\mathbf{k}_i} = \sqrt{\xi_\mathbf{k_i}^2 + |\Delta|^2},
$
and they are also extracted and shown on the right panel.
}
\end{figure}

\end{center}
\end{widetext}
\section{Numerical Results}
\label{sec:numerics} 

As we discussed thoroughly in the previous section, it is possible to have 
quantum phase transitions between SF phases with distinct $\mathbf{k}$-space 
topologies, assuming $|\Delta| \ne 0$ still pertains at sufficiently 
high Zeeman fields satisfying the condition $h = h_{\mathbf{k}_0}$.
In this section, we solve self-consistency Eqs.~(\ref{eqn:gap}), (\ref{eqn:ntot}) 
and (\ref{eqn:ndif}) at $T = 0$, and present the resultant ground-state phase 
diagrams as functions of total particle filling $F = (N_\uparrow + N_\downarrow)/M$ 
and NNN hopping $t'/t$ for various polarizations 
$
P = (N_\uparrow - N_\downarrow)/(N_\uparrow + N_\downarrow),
$
interaction strengths $g/t$ and SOC strengths $(\alpha, \beta)$.
Note that, since $0 \le N_\sigma/M \le 1$ within the single-band Hubbard model 
used in this paper, $F$ has an upper bound for a given $P$ as 
$F_\textrm{max} = 2/(1 \pm P)$ depending on $P \gtrless 0$, and that we 
assume $P \ge 0$ in this paper without loosing generality. The normal 
phase is characterized by $|\Delta| < 10^{-3} t$ in our numerical calculations.

First, we study the effects of polarization in Fig.~\ref{fig:P}, where 
we set $\alpha = \beta = \pi/4$ and $g = 10t$, and show colored maps of 
$|\Delta|/t$ and the corresponding boundaries for the topological quantum 
phase transitions as functions of $0 \le F \le 2/(1 + P)$ and $t'/t$ for 
(a) $P = 0.1$, (b) $0.3$, (c) $0.4$ and (d) $0.7$. 
The SF phase is always gapped, and therefore, it is topologically trivial 
for low Zeeman fields as long as $h < h_{\mathbf{k_0}}$ for all $\mathbf{k_0}$ 
or equivalently $P = 0$. However, the intricate dependence of DoS on $\omega$ 
and $t'$ shown in Fig.~\ref{fig:dos} gives rise to an equivalently 
intricate dependence of $|\Delta|$ on $0 \le F \le 2$ and $t'$. That is, 
not only the $D(\omega, t') = D(-\omega, -t')$ symmetry causes 
$|\Delta(F, t')| = |\Delta(2-F, -t')|$, but also $|\Delta|$ is pronounced/suppressed 
whenever $D(\omega)$ is pronounced/suppressed as intuitively expected 
from the BCS theory. For instance, in the absence of a SOC, the peak value 
of $|\Delta|$ first shifts from $F = 1$ to higher (lower) values as $t'$ increases 
(decreases) from $0$ to $0.5t$ ($-0.5 t$), and then it gradually returns 
back to $F = 1$ in the $|t'| \gg t$ limit. In the presence of a SOC, however, 
the narrow strips in $D(\omega)$ causes multiple reentrant SF phases, 
separated by the fingering normal phase, as a function of $F$.

Even though we do not present $|\Delta|$ in the strict $P = 0$ limit, remnants 
of the reentrant SF phases are clearly seen in our $P \ne 0$ figures. 
Note that since increasing $P$ dramatically weakens $|\Delta|$ for 
a given $t'/t$, additional features of $D(\omega)$ shown in Fig.~\ref{fig:dos}(c) 
vaguely appear in Fig.~\ref{fig:P}(d) but somewhat hindered by the 
dominant effects of strong $g/t$.
In addition to the direct connection between $D(\omega)$ and $|\Delta|$, 
these figures also illustrate how topological phase transition boundaries 
gradually enter the stage as $P$ increases from $0$.
Since a gapless-SF phase can be considered as a coexistence of gapped-SF 
and normal phases, we expect them to exist at the interfaces
between gapped SF and normal phases. This is best illustrated in the 
$P = 0.1$ phase diagram shown in Fig.~\ref{fig:P}(a), where the 
gapless-SF phases first emerge in the parameter regions where $|\Delta|$ is
most weakest. The gapless-SF phases progressively occupy more and 
more territory in Figs.~\ref{fig:P}(b), ~\ref{fig:P}(c) and~\ref{fig:P}(d) 
with increasing $P$.

Similarly, we study the effects of interaction strength in Fig.~\ref{fig:g}, where 
we set $\alpha = \beta = \pi/4$ and $P = 0.5$, and show colored 
maps of $|\Delta|/t$ and the corresponding boundaries for the topological 
quantum phase transitions as functions of $0 \le F \le 4/3$ and $t'/t$ for 
(a) $g = 4t$, (b) $5t$, (c) $7t$ and (d) $10t$.
Note that, since $g/t$ is relatively weak, and therefore, $|\Delta|$ is considerably 
small in Figs.~\ref{fig:g}(a) and~\ref{fig:g}(b), additional features of $D(\omega)$ 
shown in Fig.~\ref{fig:dos}(c) vaguely appear in these figures especially 
around $|t'| \sim t$, but they are again somewhat disguised in 
Figs.~\ref{fig:g}(c) and~\ref{fig:g}(d) where stronger $g/t$ effects dominate. 

Lastly, we study the effects of SOC strength in Fig.~\ref{fig:ab}, where 
we set $g = 10t$ and $P = 0.5$, and show colored 
maps of $|\Delta|/t$ and the corresponding boundaries for the topological quantum 
phase transitions as functions of $0 \le F \le 4/3$ and $t'/t$ for 
(a) $\alpha = \beta = 0$, (b) $\pi/8$, (c) $\pi/3$  and (d) $\pi/2$.
These figures further reveal the direct connection between the 
$D(\omega, t') = D(-\omega, -t')$ symmetry and that of $|\Delta(F, t')| = |\Delta(2-F, -t')|$ 
symmetry in spite of the apparent asymmetry of $|\Delta|$ caused by $P \ne 0$. 
For instance, while the reentrant structure ceases to exist and the SF 
phase occupies a compact territory in Fig.~\ref{fig:ab}(a), it can still be seen 
in Figs.~\ref{fig:ab}(b), \ref{fig:ab}(c) and~\ref{fig:ab}(d) in spite of strong $g/t$. 
We emphasize that even though the presence of a SOC is indispensable for 
creating a topological-SF phase, it is not one of the essential 
ingredients that bring about quantum phase transitions between 
SF phases with distinct $\mathbf{k}$-space topologies.
In the absence of a SOC, these transitions are solely driven by $h$ causing 
eventually $P \ne 0$, and they are clearly illustrated on the right 
panel in Fig.~\ref{fig:ab}(a).

Given these numerical results, we firmly establish that a plethora of 
quantum phase transitions are possible between SF phases with distinct 
zero-energy quasiparticle/quasihole excitation topologies in
$\mathbf{k}$ space in our model Hamiltonian given in Eq.~(\ref{eqn:ham.k}). 
Having achieved our primary objective, we are ready to end the paper with 
a brief summary of our conclusions and an outlook.

\section{Summary and Outlook}
\label{sec:conc} 

To summarize, here we considered a two-component Fermi gas with attractive 
interactions on a square optical lattice, and studied the combined effects of 
Zeeman field, SOC and NNN hopping on the ground-state phase diagrams 
in the entire BCS-BEC evolution.
For this purpose, we first discussed the effects of SOC both on the single-particle 
problem and on the corresponding DoS, and derived the self-consistency 
mean-field equations for handling the many-body problem. We then classified 
and distinguished all possible SF phases by the $\mathbf{k}$-space topology 
of their zero-energy quasiparticle/quasihole excitations, and numerically 
established that numerous quantum phase transitions are possible in 
between driven mainly by the Zeeman field. In addition, we also derived 
analytical expressions for the changes in the CN showing that these phase 
transitions are further signalled and evidenced by the changes in the 
topological invariant of the system when there is SOC. 

In addition to these important results, we found that the SF phase exhibits 
a reentrant structure, separated by a fingering normal phase, as a function
of total particle filling. This intricate structure is a result of combined SOC and 
NNN hopping, and we traced its origin back to the single-particle DoS 
of the system which is shown to exhibit a number of narrow strips (depending 
on the SOC strength) as the energy changes for a given NNN hopping. 
Given the ongoing experimental interest in spin-orbit coupled systems in 
various physics communities, even though we are aware of many technical 
difficulties in observing much of these results in the ongoing cold-atom 
experiments, we hope that some of their signatures may still be realized 
in the foreseeable future. For instance, while the presence of a SOC is 
crucial for creating both topological-SF phases and reentrant 
SF behavior, the aforementioned quantum phase transitions between SF 
phases with distinct $\mathbf{k}$-space topologies are exclusively driven 
by the Zeeman field, and therefore, they can be observed in the vanishing 
SOC and/or NNN hopping limits as well. We, at least, believe that the rich 
physics involved in this simple model may shed some light and trigger 
further research in related problems. 

As an outlook of this work, even though we expect competing SF phases 
such as the ones involving finite center-of-mass pairing to play a minor role 
in the presence of a SOC~\cite{iskin13a}, the possibility of Fulde-Ferrell-Larkin-Ovchinnikov
like SF phases, \textit{e.g.}, a topological Fulde-Ferrell SF~\cite{zhang13, qu13b, liu15}, 
may still be explored in this model. Since such a task requires a much higher 
computing power, it is beyond the scope of this paper. 
Alternatively, instead of the $\mathbf{k}$-space description employed here, 
one can use the real-space Bogoliubov-de Gennes description and
include the finite-size effects caused by the trapping potential~\cite{iskin13b}. 
The latter is a fully numerical approach and it may not only provide a way 
more complete understanding of the competing SF phases (at a cost of less 
analytical insight into their nature), but also the fate of spontaneous 
mass/spin currents, polarization/spin textures, etc. can as well be investigated.
Last but not least, our preliminary calculations on triangular lattices suggest that 
it may be valuable to extend this formalism to other kinds of two-dimensional 
geometries including the honeycomb lattice, an important direction that is highly 
feasible in the advent of tunable optical lattices~\cite{jotzu14, uehlinger13}.

\section{Acknowledgments} 
\label{sec:ack}
We gratefully acknowledge funding from T\"{U}B$\dot{\mathrm{I}}$TAK 
Grant No. 1001-114F232.

\end{document}